\documentclass[aps,twocolumn,amsmath,amssymb,showpacs]{revtex4}
\usepackage{epsf}
\newcommand{\be}{\begin{equation}}
\newcommand{\ee}{\end{equation}}
\newcommand{\bea}{\begin{eqnarray}}
\newcommand{\eea}{\end{eqnarray}}

\newcommand{\sign}{{\rm sign}\, }
\begin{document}

\title {First order superconducting transition near a ferromagnetic
quantum critical point}

\author{Andrey V. Chubukov$^1$, Alexander M. Finkel'stein$^2$,
 Robert Haslinger$^3$,  and Dirk K. Morr$^4$}
\affiliation{ $^1$ Department of Physics, University of Wisconsin,
Madison, WI 53706\\
$^2$ Department of Physics, Weizmann Institute of Sciences,
Rehovot, Israel\\
$^3$ Los Alamos National Laboratory, Los Alamos, NM 87545 \\
$^4$ Department of Physics, University of Illinois at Chicago,
Chicago IL 60607}

\date{\today}
\begin{abstract}
We address the issue of how triplet superconductivity emerges in
an electronic system near a ferromagnetic quantum critical point
(FQCP). Previous studies found that the superconducting transition
is of second order, and $T_c$ is strongly reduced near the FQCP
due to pair-breaking effects from thermal spin fluctuations. In
contrast, we demonstrate that near the FQCP, the system avoids
pair-breaking effects by undergoing a {\it first order} transition
at a much larger $T_c$. A second order superconducting transition
emerges only at some distance from the FQCP.
\end{abstract}

\pacs{PACS numbers: 74.25.-q}

\maketitle

Superconductivity near a magnetic instability is a topic of
current interest in condensed-matter physics. Magnetically
mediated pairing near an antiferromagnetic instability is a
candidate scenario for $d-$wave  superconductivity in the cuprates
and heavy fermions compounds (for a recent review see
\cite{pines}). The emergence of superconductivity, mediated by the
exchange of ferromagnetic spin fluctuations, is also expected near
ferromagnetic transitions. Ferromagnetic exchange yields Cooper
pairs with $S=1$ and therefore generally gives rise to triplet
superconductivity. This type of pairing was originally suggested
by Anderson and Morel \cite{am} for $^{3}He$. In recent years, an
intensive search has focused on superconductivity in compounds
which can be tuned to a ferromagnetic quantum critical point
(FQCP) by varying either pressure or chemical composition. Among
the studied systems are MnSi, and the heavy fermion compound
UGe$_2$ (for an experimental review, see Ref.~\cite{fm-hf}).

The emergence of superconductivity in electronic systems close to
a ferromagnetic instability has recently been studied by three
groups, who solved a linearized gap equation within the Eliashberg
formalism \cite{lonzarich,bedell,millis}. For both two- (2D) and
three-dimensional (3D) systems, their analysis yielded a
superconducting transition temperature, $T^{l}_{c}$ ('l' stands
for linearized), that substantially decreases as the system
approaches criticality, eventually vanishing at the FQCP.

The physical origin of the decrease in $T^{l}_{c}$ near the FQCP
lies in the presence of thermal spin fluctuations which behave
like magnetic impurities whose scattering potential diverges as
the critical point is approached \cite{lonzarich,bedell,millis}.
This behavior is reflected in the fermionic self-energy in the
normal state, $\Sigma (\omega_n) \propto i T \sum_m
\sign(\omega_n) \chi_L (\omega_m -\omega_n)$, where $\chi_L
(\omega) = \int d^{D-1} q \chi (q, \omega)$ is the ``local'' spin
susceptibility. Since $\chi_L (\omega=0)$ diverges at the FQCP for
$D \leq 3$ (assuming $\chi (q, 0)=\chi_0/(\xi^{-2}+ q^2)$ with
$\xi \rightarrow \infty$), the dominant contribution to
$\Sigma(\omega_n)$ comes from the $n=m$ term in the frequency sum,
i.e., from classical, thermal spin fluctuations. These
fluctuations scatter with a finite momentum and zero frequency
transfer; hence their similarity with magnetic impurities. This
leads to $\Sigma (\omega) = i \gamma \sign \omega$ with $\gamma
\propto T\xi ^{3-D}$ for system dimension $D<3$ and $\gamma
\propto T\log \xi $ for $D=3$.

The analogy with magnetic impurities extends to the pairing
problem in which thermal spin fluctuations close to the FQCP tend
to break Cooper pairs and hence lower the temperature of the
superconducting transition\cite{ag_2}. This strong pair breaking
effect is reflected in the gap equation where the divergent
contributions to the self-energy $\Sigma (\omega)$ and to the
pairing vertex are {\it not} cancelled out in the equation for the
pairing gap $\Delta (\omega)$ \cite{Lit92} (the ratio of divergent
terms is 3 to 1 in our case). Simple estimates show that the
linearized gap equation does not have a solution above $\gamma
\sim T_{c0}$ where $T_{c0}$ is the transition temperature in the
absence of thermal fluctuations. Hence, when the FQCP is
approached, $T_{c}^{l}$ vanishes as $T_{c}^{l}\propto \xi ^{D-3}$
in $D<3$, and $T_{c}^{l}\propto 1/\log \xi $ for $D=3$. Numerical
solutions of the linearized Eliashberg equations near a
ferromagnetic instability demonstrate precisely this kind of
behavior - $T_{c}^{l}$ falls off when the FQCP is approached -
more rapidly in $2D$~\cite{lonzarich} than in $3D$~ \cite
{bedell,millis} (Refs.~\cite{bedell,millis} obtained a small
finite $T^l_{c}$ at criticality by using a self-consistent
approach that goes beyond Eliashberg theory for the self-energy,
but still neglects vertex corrections).

In this letter, we argue that the actual behavior of the system is
different from that discussed in
Refs.~\cite{lonzarich,bedell,millis}. Specifically, we show that
close to a FQCP, superconductivity emerges via a {\it first order
phase transition} at $T_c \sim T_{c0}$. The much smaller
$T_{c}^{l}$ previously obtained by solving the linearized gap
equation is just the end point of the temperature hysteresis loop,
at which the normal state becomes unstable. The first indication
that the pairing problem near the FQCP is unconventional comes
from the observation that at $T=0$, dangerous thermal fluctuations
are absent, and hence the pairing gap $\Delta (\omega)$ should
generally be of the order of $T_{c0}$. Explicit calculations
confirm this (see below). The second indication is that in the
presence of a large gap, the spectrum of ferromagnetic spin
fluctuations changes due to feedback effects from the pairing. For
the ABM phase of triplet superconductivity, which we consider in
the following, this feedback is different for $\chi_{zz}$ and
$\chi_{\pm}$ \cite{am} (assuming that the spin of the Cooper pair,
${\vec S}_{cp}$, lies in the xy-plane). At the FQCP, massless
excitations survive in $\chi_{\pm}$, but not in $\chi_{zz}$ which
in the presence of a pairing gap describes massive longitudinal
spin fluctuations. Due to this distinction between $\chi_{\pm}$
and $\chi_{zz}$, the ratio of the divergent terms in the
self-energy and the pairing vertex is $\left[ 2 \chi^L_{\pm} (0) +
\chi^L_{zz} (0)\right]/\left[ 2 \chi^L_{\pm} (0) - \chi^L_{zz}
(0)\right]$. If $\chi^L_{\pm}$ and $\chi^L_{zz}$ behaved
identically, this ratio would be $3$, and pair-breaking effects of
thermal fluctuations would be crucial. When only $\chi^L_{\pm}(0)$
diverges, the above ratio is $1$ and the divergent terms from the
fermionic self-energy and the pairing vertex cancel out in the gap
equation. This in turn implies that the superconducting state with
a large gap remains stable well above $T^{l}_{c}$. This behavior,
however, cannot extend to small $\Delta (\omega)$ since then
$\chi^L_{zz} (0)$ cannot be neglected, the above ratio of
divergent terms becomes $3$ and no cancellation occurs. We
therefore expect that at the FQCP, the solution with a finite gap
should survive up to the end point at $T_{c}^{nl}\sim T_{c0}\gg
T_{c}^{l}$ ('nl' stands for the solution of the nonlinear
equation) where it becomes unstable. In other words, over some
range of $T$, both the normal state and the state with a large gap
are locally stable. This is a classic scenario for a first order
transition. Note, that this behavior is very different from that
for singlet pairing, mediated by antiferromagnetic spin
fluctuations, where the divergent contributions to $\Sigma
(\omega)$ and to the pairing vertex cancel each other even in the
linearized gap equation. As a result, the divergence of the
thermal self-energy at criticality does not affect $T^{l}_{c}$
\cite{acf} which saturates at a finite value for $\xi = \infty$.

In the remainder of this paper we compute $T^{nl}_{c}$ from the
full set of nonlinear Eliashberg equations and show that it
saturates at a finite value at the FQCP. We will not attempt to
compute the actual $T_c$ (this would require the analysis of the
condensation energy). However, since $T^l_{c}$ vanishes at  the
FQCP and $T^{nl}_{c}$ stays finite, $T_c$ near criticality should
be comparable to $T^{nl}_{c}$.

Our starting point is the spin-fermion model, which describes the
interaction of low-energy fermions with their own spin degrees of
freedom, $\mathbf{S}_{\mathbf{q}}$, whose propagator is peaked at
$q=0$. The same model was used in earlier studies
\cite{abanov_review,lonzarich,bedell,millis,acf}. We assume that
$T_{c0}$ is much less than the Fermi energy $E_{F}$, implying that
the pairing instability involves only fermions near the Fermi
surface. The model is described by the Hamiltonian
\begin{eqnarray}
\mathcal{H}&=&\sum_{\mathbf{k},\alpha }\mathbf{v_{F}}(\mathbf{k}-\mathbf{k}%
_{F})c_{\mathbf{k},\alpha }^{\dagger }c_{\mathbf{k},\alpha
}+\sum_{q}\chi
^{-1}(\mathbf{q})\mathbf{S}_{\mathbf{q}}\mathbf{S}_{-\mathbf{q}} \nonumber \\
&&+g\sum_{%
\mathbf{q,k},\alpha ,\beta }~c_{\mathbf{k+q},\alpha }^{\dagger }\,\mathbf{%
\sigma }_{\alpha ,\beta }\,c_{\mathbf{k},\beta }\cdot \mathbf{S}_{\mathbf{-q}%
}\,. \label{intham}
\end{eqnarray}
where the spin-fermion coupling, $g$, the  Fermi velocity, $v_{F}$
(we assume a circular Fermi surface), and the static spin
propagator $\chi (q,0)$ are input parameters. While the upper
energy cutoff is in general also an input parameter, our results
are cutoff independent for the pairing problem considered here.
The dynamical part of $\chi (q,\Omega_{m})=\chi _{0}/[q^{2}+\xi
^{-2}+\Pi (q,\Omega _{m})]$ arises from the interaction with the
low-energy fermions and is explicitly calculated. While we
restrict our consideration to $D=2$, our conclusions are also
valid for $3D$ systems.

We assume that the static $\chi (q,0)$ has the conventional
lorentzian form  with a weakly temperature dependent $\xi$. This
form of $\chi(q,0)$ was recently questioned~\cite{bkv,millis2}
since far away from criticality, the static spin susceptibility
possesses singular low-energy {\it Fermi liquid} corrections that
give rise to a universal $|q|$ dependence of $\chi (q,0)$, and a
$T$ dependence of $\xi^{-1}$. It is unclear, however, whether
these singular corrections survive in the quantum critical regime,
so we restrict with the conventional form of $\chi(q,0)$ without
further justifications.

Near the critical point, a conventional perturbation theory in the
spin-fermion coupling (for which $\Sigma = \Pi =0$ is the point of
departure) holds in powers of $\lambda =g^{2}\chi _{0}/(4\pi
v_{F}\xi ^{-1})$, i.e., the quantum-critical region falls into the
strong coupling limit. An approach for dealing with a strong
coupling problem is the Eliashberg theory~\cite{eliash}. Its
validity requires certain conditions to be met. We proceed
assuming that the Eliashberg theory is valid, and then discuss
what restrictions are necessary.

We first consider the situation right at the FQCP where $\xi^{-1}
=0$. In the normal state, the dynamical part of the spin
polarization operator, $\Pi(q,\Omega _{m})$, is independent of the
fermionic self-energy, $\Sigma(q,\omega _{n})$ (but not vice
versa), as the essential momenta for $\Pi(q,\Omega_{m})$ are those
with $v_{F}q \gg \Omega_m, \Sigma$. As a result, $\Pi (q,
\Omega_m)$ has the same form as for free fermions, i.e., for
$|\Omega_m| \ll v_F q$, $\Pi (q, \Omega_m) = {F} (\Omega_m)/(v_F
q)$ where ${F} (\Omega_m) = \alpha k^2_F |\Omega _{m}|$, $\alpha =
g^2 \chi_0/(2\pi E_F)$, and $E_F = k_F v_F/2$. At the same time,
$\Sigma (\omega)$ is determined by $\Pi$ and given by $\Sigma
(\omega_m) = \,\omega _{0}^{1/3}\,\omega _{m}^{2/3}$, where
$\omega_0 = (3 \sqrt{3}/4) \alpha^2 E_F$. The non-Fermi liquid,
$\omega^{2/3}$-dependence of the self-energy in 2D is due to the
divergence of the perturbation theory at the FQCP. This form was
earlier obtained in Ref.~\cite{aim}.

In the superconducting state, the equations for two components of
$F(\Omega)$ (${F}_{zz}$ and ${F}_{xx} = {F}_{yy}$) are coupled to
the equation for the pairing gap $\Delta (\omega)$. Similar to the
normal state, none of these quantities explicitly depends on
$\Sigma$. As a result, one needs to self-consistently solve a set
of three coupled equations for the two components of ${F}$ and
$\Delta$. The derivation of the Eliashberg equations is quite
straightforward and will not be presented here. With the Cooper
pair spin lying in the xy-plane, the  coupled equations  for
$\Delta (\omega)$, ${F}_{zz} (\Omega) ={F}_- $ and ${F}_{xx} =
{F}_{yy} = {F}_+$ at the FQCP have the form
\begin{eqnarray}
&&\Delta \,\left( \omega _{n}\right) =\frac{4\pi }{9}
\omega^{1/3}_0~ T\sum_{m}\frac{1}{%
\sqrt{\omega_{m}^{2}+\Delta ^{2}(\omega _{m})}}%
 \nonumber \\
&& \times \left\{ \frac{2 \omega_m}{\left[ {F}_{+}(T,\Delta ,\omega_m -\omega _{n})\right] ^{1/3}}%
\left[\frac{\Delta (\omega _{m})}{\omega_m} -\frac{\Delta \,\left( \omega
_{n}\right) }{\omega _{n}} \right]
\right. \nonumber\\
&& \hspace{-0.3cm} \left. -\frac{\omega_m}{\left[ {F}_{-}(T,\Delta
,\omega_m -\omega _{n})\right] ^{1/3}}\left[ \frac{\Delta (\omega
_{m})}{\omega_m} +\frac{\Delta \,\left(
\omega _{n}\right) }{\omega _{n}} %
\right] \right\}
\label{delta}
\end{eqnarray}
and
\begin{eqnarray}
&&{F}_{\pm }(T,\Delta ,\omega_m - \omega_n)= \nonumber \\ & & \pi
T\sum_{n} \left[ 1- \frac{\omega _{n} (\omega _m)\pm \Delta
(\omega _{n})\Delta (\omega _m)}{\sqrt{\omega _{n}^{2}+\Delta
^{2}(\omega _{n})} \sqrt{\omega _{m} ^{2}+\Delta ^{2}(\omega
_{m})}}\right] \label{F}
\end{eqnarray}
As anticipated, ${F}_+ (T,\Delta ,0) =0$ vanishes implying that
the corresponding susceptibilities, $\chi_{xx}$ and $\chi_{yy}$,
describe massless modes. The vanishing of ${F}_+ (T,\Delta ,0)$,
however, is not dangerous as it is compensated by the simultaneous
vanishing of the numerator in Eq.(\ref{delta}). Exactly the same
cancellation of divergences occurs in the gap equation for
$d$-wave pairing due to antiferromagnetic spin fluctuations. The
vanishing of  ${F}_- (T,\Delta ,0)$ would be dangerous, but for
finite $\Delta$,  ${F}_- (T,\Delta ,0) = 2 \pi T \sum_n
\Delta^2(\omega_n)/\sqrt{\omega^2_n + \Delta^2(\omega_n)}$ is also
finite, i.e. the longitudinal spin excitation described by
$\chi^{-1}_{zz}(\omega=0) \propto q^2 + {F}_- (T,\Delta ,0)/(v_F
q)$ is massive.  In contrast, for the linearized gap
equation $\Delta$ is vanishingly small, ${F}_- (T,\Delta,0)$
vanishes, and the r.h.s of Eq.(\ref{delta}) diverges. Due to this
divergence, the linearized gap equation does not have a solution
down to $T=0$. Note, that the only energy scale in the
Eliashberg equations is $\omega_0$, which can be eliminated by
rescaling both temperature and the gap in units of $\omega_0$. The
gap equation is then {\it fully universal}, which implies that the
mass in $\chi_{zz}$ and the typical momentum $q$ for the pairing
problem are both of order $\omega_0/v_F$.

\begin{figure}[tbp]
\begin{center}
\epsfxsize=\columnwidth \epsffile{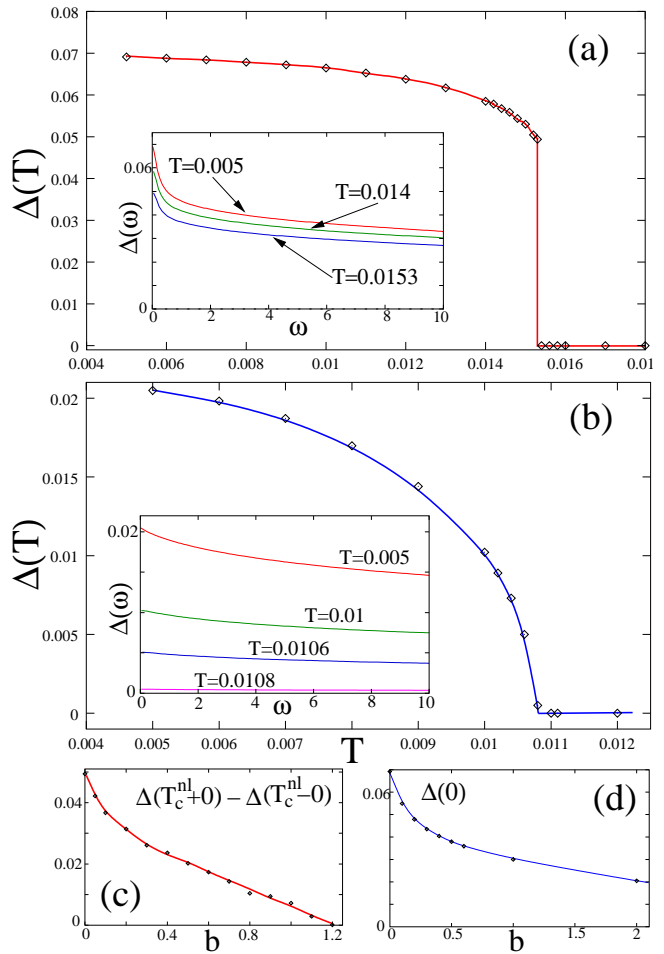}
\end{center}
\vspace{-0.75cm} \caption{(a) Temperature dependence of $\Delta
(T,\omega_m)$ at the lowest Matsubara frequency $\omega_m = \pi T$
for  $b=0$. The lines are a guide for the eye. $\Delta$, $T$ and
$\omega_m$ are in units of $\omega_0$ (see text).  The
discontinuity of $\Delta (T)$ at $0.015$ indicates a first order
transition. The inset shows $\Delta (\omega)$  versus frequency at
several $T$. (b) Same away from the FQCP, for $b=2$. Now the
transition is continuous. (c)  The magnitude of the jump of
$\Delta (T, i\pi T)$ at the instability temperature versus $b$.
The line is a guide to the eye. (d) $\Delta(T, i\pi T))$ at the
lowest $T$ versus $b$.} \label{Fig1}
\end{figure}
In Fig.~\ref{Fig1}a we present the numerical solution for $\Delta
(T)$ at the lowest Matsubara frequency, $\omega_m=\pi T$. As
expected for a first order phase transition, the gap changes
discontinuously from a finite value to zero at $T^{nl}_{c} \sim
0.015 \omega_0$. The inset shows that the discontinuous jump in
$\Delta$ occurs for all Matsubara frequencies.

We next study the situation at finite $\xi$ and verify whether the
first order superconducting transition becomes second order at
some distance from the FQCP. Away from criticality, the equations
for ${F}_{\pm} (T,\Delta ,\omega_m)$ retain their form, but in the
gap equation, the factors ${F}_\pm (T,\Delta ,\omega_m)$ are
replaced by ${F}_{\pm} (T,\Delta ,\omega_m)/I^3(\beta_{\pm})$,
where $I(\beta_{\pm})$ by itself depends on ${F}_\pm (T,\Delta
,\omega_m)$ through
\begin{equation}
I(\beta_{\pm})=\frac{3\sqrt{3}}{2\pi}~\int_{0}^{\infty }dx
\frac{x}{1+\beta_{\pm }x+x^{3}} \ ,
\end{equation}
Here $\beta_{\pm } = b/[{F}_{\pm}(T,\Delta ,\Omega _{m})/\omega
_{0}] ^{2/3}$, and $b = (8/3\sqrt{3})^{2/3} (\alpha k_F \xi)^{-2}$
measures the deviation from the FQCP ($b=0$ at the FQCP). The
integral can be expressed in elementary functions. One can easily
verify that for finite $\xi$, the gap equation does not contain
any divergence, even at infinitesimally small $\Delta$, and hence
both $T^{nl}_{c}$ and $T^{l}_{c}$ are nonzero.

In Fig.~\ref{Fig1}b we plot $\Delta (T, i\pi T)$ for $b=2$. We
clearly see that the temperature dependence of the gap is now
continuous, in marked distinction to Fig.~\ref{Fig1}a. The inset
shows that the continuous evolution of $\Delta$ holds for all
Matsubara frequencies. This result implies that for large enough
$b > b_c$, the transition is of second order. To locate the
tricritical point $b = b_c$, we plot in Fig. \ref{Fig1}c the
magnitude of the jump of $\Delta (T, i\pi T)$ at $T^{nl}_{c}$ as a
function of $b$.  We see that the gap discontinuity disappears at
$b_c \approx 1.2$. For completeness, in Fig.~\ref{Fig1}d we plot
the zero temperature gap versus $b$. We see that it changes
gradually, without a singularity at $b_c$.

\begin{figure}[t]
\begin{center}
\epsfxsize=\columnwidth \epsffile{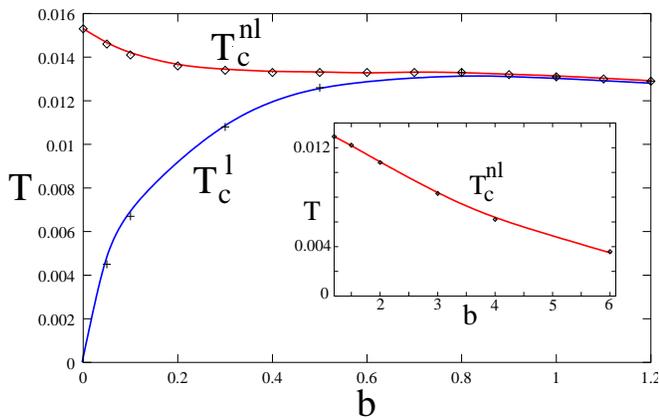}
\end{center}
\vspace{-0.75cm} \caption{The phase diagram near the FQCP. In the
near vicinity of the FQCP, the transition is of first order, away
from the FQCP, to the right of $b_c$, it is of second order. For
the first order transition, $T^{nl}_{c}$ and $T_c^{l}$ are the
instability temperatures for the solutions with a large and
infinitesimally small gap, respectively. The actual first order
transition temperature, $T_c$, lies between $T^l_{c}$ and
$T^{nl}_{c}$. Inset: the reduction of $T_c$ at large $b$.}
\label{Fig2}
\end{figure}

The phase diagram that emerges from our studies is presented in
Fig.~\ref{Fig2}. The actual superconducting transition temperature
lies between $T^{l}_{c}$ and $T^{nl}_{c}$ and hence remains {\it
finite} at $b =0$. For $0.9 \lesssim b<b_c$, the jump in $\Delta$
at $T^{nl}_{c}$, and consequently the difference between
$T^{l}_{c}$ and $T^{nl}_{c}$ is small but finite (see
Fig.~\ref{Fig1}c). The insert shows the reduction of $T_c$ at
large $b$ due to the decrease in the effective coupling.

The functional form of $T^{l}_{c}(b)$ obtained above agrees
qualitatively  with that found earlier by Monthoux and Lonzarich
\cite{lonzarich}. Their result for the maximum $T^l_{c}$ is
$T^l_{c,max} \sim 0.016 \, \omega_0 \,b^{3/2}$. Substituting $b
\sim 0.8$, where $T^{l}_c$ is maximal in our case, we obtain
$T^l_{c,max} \sim 0.011 \omega_0$, consistent with our result
$T^l_{c,max} \sim 0.013 \omega_0$.

Finally, the validity of the Eliashberg treatment requires that
three conditions be satisfied. First, typical bosonic momenta
$q_B$ should be much larger than typical fermionic $|k-k_F|$,
i.e., bosons should be slow modes compared to fermions, leading to
$\Sigma (k, \omega) \approx \Sigma (\omega)$. A straightforward
analysis shows that the typical $q_B \sim \alpha k_F$, while the
typical $|k-k_F| \sim \omega_0/v_F \sim \alpha^2 k_F$. The
Eliashberg theory is therefore valid if $\alpha \ll 1$, i.e.,
$\omega_0 \ll E_F$. This in turn implies that the physical
behavior of the system is universally determined by only low
energy excitations. Second, vertex corrections should be small.
Generally, this is not possible for typical $q_B \ll k_F$, as
vertex corrections scale with $\xi$. However, for $\alpha \ll 1$,
we only require the vertex for $\Omega \ll v_F q$ since the
typical $v_F q_B$ well exceeds the typical $\Omega \sim \omega_0$.
In this limit, vertex corrections are much smaller and only scale
as $\log \xi$. They are  still non-negligible at the FQCP, where
they change the pole in the spin susceptibility into a branch cut
\cite{abanov_review}. However, we verified that, as in the
antiferromagnetic case \cite{acf}, this only leads to a small
renormalization of the prefactors in the gap equation. Third, one
should be able to neglect the momentum dependence of the pairing
gap, while preserving the gap symmetry, $\Delta ({\vec n} k_F) = -
\Delta (-{\vec n} k_F)$. This approximation is again justified by
$\alpha \ll 1$, as in this limit, the typical momentum transfers
along the Fermi surface $\delta k \sim q_B \sim \alpha k_{F}$ are
much smaller than $k_{F}$. In this situation, the momentum
variation of the gap at typical $\delta k$ only introduces $O(1)$
corrections to the Eliashberg theory \cite{acf}, which can be
safely neglected.  Note in passing that the smallness of  $\delta
k \ll k_F$ makes our theory also applicable to real materials (in
which a crystalline structure imposes additional constraints on
the order parameter symmetry~\cite{VG}), as it allows one to
consider the pairing problem in a local-momentum approximation,
ignoring the peculiarities of the gap's momentum dependence.

In summary, we showed that near a FQCP, spin fluctuation exchange
gives rise to a strong first order transition into a $p-$wave
superconducting state. By choosing a first order transition, the
system avoids divergent pair-breaking effects from thermal spin
fluctuations. As a result, $T_c$ saturates at a nonzero value at
criticality. The first order transition persists up to a finite
distance from the FQCP, where it becomes second order.

It is our pleasure to thank A. Abanov, D. Khveschenko, D. Pines,
A. Tsvelik and Z. Wang for useful discussions. The research was
supported by NSF DMR-9979749 (A. Ch.), BSF-1999354 (A.C and A. F)
and by DR Project 200153 and the Department of Energy, under
contract W-7405-ENG-36. (R.H.) A.C. and D.M. would like to thank
Los Alamos National Laboratory for its hospitality during the
completion of this project. 

\end{document}